\documentclass[aps,twocolumn,prc,showpacs,preprintnumbers,
               nofootinbib,float,superscriptaddress,longbibliography]{revtex4-1}
\usepackage{dcolumn}
\usepackage{bm}
\usepackage{amssymb}
\usepackage{float}
\usepackage[caption = false]{subfig}

\usepackage{footnote}

\usepackage{hyperref}
\usepackage{xcolor}
\usepackage{graphicx}

\usepackage{epsfig}
\usepackage{color}
\usepackage{lineno}

\newcommand{\p}{p(p)}
\newcommand{\de}{d(pn)}
\newcommand{\tr}{t(pnn)}
\newcommand{\lam}{\Lambda(\Lambda)}
\newcommand{\hethree}{{}^{3}_{}\mathrm{He}(ppn)}
\newcommand{\hefour}{{}^{4}_{}\mathrm{He}(ppnn)}
\newcommand{\hyphthree}{{}^{3}_{\Lambda}\mathrm{H}(pn\Lambda)}
\newcommand{\hyphfour}{{}^{4}_{\Lambda}\mathrm{H}(pnn\Lambda)}

\usepackage{tikz}
\usetikzlibrary{shapes.geometric, arrows}

\tikzstyle{startstop} = [rectangle, rounded corners, minimum width=3cm, minimum height=1cm,text centered, draw=black, fill=red!30]
\tikzstyle{io} = [trapezium, trapezium left angle=70, trapezium right angle=110, minimum width=3cm, minimum height=1cm, text centered, draw=black, fill=blue!30]
\tikzstyle{process} = [rectangle, minimum width=3cm, minimum height=1cm, text centered, draw=black, fill=orange!30]
\tikzstyle{decision} = [rectangle, minimum width=3cm, minimum height=1cm, text centered, draw=black, fill=green!30]
\tikzstyle{comment} = [rectangle, rounded corners, minimum width=3cm, minimum height=1cm,text left, text width=4.2cm, draw=black, fill=yellow!30]
\tikzstyle{arrow} = [thick,->,>=stealth]

\begin{document}
\title{
 Investigating the coalescence-inspired sum rule for light nuclei and hypernuclei in heavy-ion collisions
}

\author{\it Ashik Ikbal Sheikh}
\email{asheikh2@kent.edu / ashikhep@gmail.com}
\address {Department of Physics, Kent State University, Kent, OH 44242, USA}


\date{\today}

\begin{abstract}

 A data-driven idea is presented to test if light nuclei and hypernuclei obey the coalescence-inspired sum rule, i.e., to test if the flow of a light nucleus or hypernucleus is the summed flow of each of its constituents. Here, the mass difference and charge difference among the constituents of light nuclei and hypernuclei are treated appropriately. The idea is applied to the available data for $\sqrt{s_{NN}} = 3$ GeV fixed-target Au+Au collisions at the Relativistic Heavy Ion Collider (RHIC), published by the STAR collaboration.  It is found that the sum rule for light nuclei is approximately valid near mid-rapidity ($-0.3 < y < 0$), but there is a clear violation of the sum rule at large rapidity ($y < -0.3$). The Jet AA Microscopic Transport Model (JAM), with baryonic mean-field plus nucleon coalescence, generates a similar pattern as obtained from the experimental data. In the present approach, the rapidity dependence of directed flow of the hypernuclei $\mathrm{_{\Lambda}^{3}H}$ and $\mathrm{_{\Lambda}^{4}H}$ is predicted in a model-independent way for $\sqrt{s_{NN}} = 3$ GeV Au+Au collisions, which will be explored by ongoing and future measurements from STAR.

\end{abstract}




\maketitle
\section{Introduction}

Collectivity is a phenomenon observed over a wide range of collision energies for various heavy-ion collision systems. The azimuthal anisotropy of emitted particles is characterized by Fourier decomposition of final-state particle momenta with respect to the reaction plane~\cite{Voloshin:1994mz,Poskanzer:1998yz}. The first and higher-order coefficients of the azimuthal anisotropy, also known as directed flow ($v_1$), anisotropic flow ($v_2$), and so on, describe a collective motion of particles. The azimuthal anisotropies provide important information
on the collective hydrodynamic expansion and transport properties of the matter formed in the collisions. They are also sensitive to the compressibility of the nuclear matter and the nuclear equation of state at collision energies of the order of a few GeV~\cite{Heinz:2013th,Danielewicz:2002pu}. The anisotropic flow coeffiecients of different identified particle species have been measured extensively in experiments at RHIC~\cite{STAR:2005btp,STAR:2008jgm,Adamczyk:2011aa,Adamczyk:2014ipa,Adamczyk:2017nxg} and the LHC~\cite{ALICE:2010suc,Acharya:2019ijj}.

Apart from the identified particles, the measurements of hypernuclei ($\mathrm{_{\Lambda}^{3}H}$, $\mathrm{_{\Lambda}^{4}H}$)~\cite{STAR:2021orx,STAR:2019wjm,STAR:2017gxa,ALICE:2015oer,STAR:2010gyg} and measured azimuthal anisotropies for light nuclei ($d$, $t$, $^3$He, $^4$He)~\cite{ALICE:2020chv,ALICE:2017nuf,EOS:1994kku,E877:1997,FOPI:2004hyz,FOPI:2011aa,STAR:2020hya,STAR:2016ydv,HADES:2020lob,ALICE:2017xrp} have also been reported in the past. Hypernuclei are natural hyperon-baryon correlation systems, and can serve as an excellent probe of hyperon-baryon interactions in high-energy heavy-ion collisions. Measurements of hypernuclei produced in the collisions have lately been of increasing interest. On the other hand, at lower collision energies, a larger anisotropic flow is measured for light nuclei compared to protons~\cite{EOS:1994kku,EOS:1994jzn,E877:1997,FOPI:2004hyz,FOPI:2011aa,STAR:2020hya,STAR:2016ydv,HADES:2020lob,STAR:2021ozh}, suggesting 
possible advantages of studying light nuclei. The STAR collaboration reported the scaling of light nuclear elliptic flow according to nuclear mass number ($A$), in a reduced transverse momentum ($p_T$) range $p_T/A<1.5$ GeV/$c$ over a wide range of collision energies, $\sqrt{s_{NN}} = 7.7 -\- 200$ GeV~\cite{STAR:2016ydv}. This observation favors the interpretation that the light nuclei are formed at these energies and kinematics via coalescence of nucleons. However, the true production mechanism of light nuclei and hypernuclei is not yet fully understood and remains 
under active research~\cite{Andronic:2017pug,Vovchenko:2016ebv,Andronic:2010qu,Zhao:2018lyf,Oliinychenko:2018ugs}. In the coalescence mechanism, light nuclei or hypernuclei are formed by the binding of nucleons or hyperons when they come close to each other in both coordinate and momentum space during the time of kinetic freeze-out~\cite{Butler:1961pr,Sato:1981ez,Zhang:2009ba}. 
The interaction between the produced expanding fireball and the spectator remnants
becomes more significant at lower beam energies due to the longer passing time of the colliding ions. The flow signals are strongly affected by the relatively slowly-passing spectators, and hence one might get important insights into the collision dynamics and the nucleon coalescence behavior. Recently, the STAR collaboration has observed a breakdown of $A$ scaling for flow of light nuclei away from mid-rapidity in $\sqrt{s_{NN}} = 3$ GeV Au+Au collisions~\cite{STAR:2021ozh}. 

In the traditional $A$ scaling for light nuclei and hypernuclei (e.g., Ref.~\cite{STAR:2021ozh}), each constituent nucleon or hyperon is on equal footing, which ignores the fact that the constituents have different masses and electric charges, whereas the resulting flow of nuclei through coalescence mechanism depends on the mass and charge of the constituents. The mass difference between proton and neutron may be negligibly small, but due to the charge difference, the Coulomb effect must be larger than the mass effect. In this article, a novel data-driven method is discussed, which tests the coalescence-inspired sum rule for light nuclei and hypernuclei, considering different constituents according to their mass and charge. 

It is hard to measure each and every constituent of a light nucleus or hypernucleus in an experiment. Hence, the idea is to combine different light nuclei and hypernuclei, then compare the combinations so that they have identical constituents, i.e., the combinations are compared at the same mass and same charge at the constituent level. The method is discussed in detail in the next section (Sec.~\ref{method}). Under this method, the sum rule is tested using the STAR measurements available for light nuclei from $\sqrt{s_{NN}} = 3$ GeV Au+Au collisions. 
A nuclear transport model named the Jet AA Microscopic Transport Model (JAM)~\cite{Nara:1999dz} with a baryonic mean field~\cite{Isse:2005nk} plus nucleon coalescence calculations is found to be quite successful in describing the measured $v_1$ and $v_2$ for light nuclei from $\sqrt{s_{NN}} = 3$ GeV Au+Au collisions~\cite{STAR:2021ozh}. The sum rule has also been tested with the same JAM model and the calculations agree with the results obtained from the STAR data at $\sqrt{s_{NN}} = 3$ GeV Au+Au collisions. 


The data-driven method predicts the rapidity dependence of $v_1$ for hypernuclei like $\mathrm{_{\Lambda}^{3}H}$ and $\mathrm{_{\Lambda}^{4}H}$ in $\sqrt{s_{NN}} = 3$ GeV Au+Au collisions. STAR has collected large data sets at various beam energies, both in fixed target and collider modes as part of Phase II of the Beam Energy Scan program~\cite{StarBur:2021}, and these detailed measurements will serve as a good testing ground for the analysis proposed in the present work.  

In the next section, details of the method are outlined. Results are discussed in Sec.~\ref{results}. Section~\ref{summary} presents a summary.

\section{Method}
\label{method}
\subsection{Coalescence-inspired sum rule in a data-driven approach}

In the proposed approach, it is assumed that light nuclei and hypernuclei are predominantly formed via coalescence of the constituent nucleons or $\Lambda$ hyperons, and it is also assumed that the anisotropic flow correlation is imposed before hadronization~\cite{STAR:2016ydv,STAR:2021orx}, i.e., well before formation of the nuclei under consideration. The abundantly produced light nuclei and hypernuclei reported by experimental collaborations to date are: $\de$, $\tr$, $\hethree$, $\hefour$, $\hyphthree$, and $\hyphfour$. The $A$ scaling for light nuclei and hypernuclei follows from the coalescence mechanism. The different constituents
of light nuclei and hypernuclei in this scaling behavior are treated equally, which ignores the fact that in general, the constituents have different masses, charges and strangeness. 
 In the following method, the coalescence-inspired sum rule for light nuclei and hypernuclei can be tested where the constituents are 
 considered depending upon their masses, charges and strangeness, i.e., the method does not ignore the mass difference, charge difference and strangeness difference of the constituents. A similar approach was developed in earlier work~\cite{Sheikh:2021rew} which focused on hadron formation via coalescence in heavy-ion collisions. 

 The first step in the present method is to select a kinematic region where the aforementioned assumptions of the sum rule can be tested, which involves a test of the equality
\begin{eqnarray}
v_1({\rm light~(hyper)nucleus}) = \sum\limits_i v_1(N_i),
\label{eq_csr}
\end{eqnarray}
where the sum runs over the $v_1$ for the nucleon or $\Lambda$ hyperon constituents, $N_i$.

The next step of the method is to combine different light nuclei and hypernuclei, then compare the combinations which have identical constituents, i.e., the combinations being  compared have the same mass and same charge at the constituent level. For example, $\p+\de$ has the identical constituent nucleons as $\hethree$. Therefore, the consistency of the sum rule can be investigated experimentally by testing the equality

\begin{eqnarray}
v_1[\p] + v_1[\de] = v_1[\hethree].
\label{eq_csr1}
\end{eqnarray}
Here, both left and right sides have the identical constituent nucleon content of $ppn$. Hence at the constituent level, the mass difference, the charge difference and the mass number difference between left and right sides are $\Delta m = 0$, $\Delta q = 0$ and $\Delta A = 0$, respectively. However, the three nucleons here are distributed differently within the two light nuclei on the left side. For convenience of discussion, such combinations are expressed in terms of a difference, $\Delta v_1$. For example, Eq.~(\ref{eq_csr1}) can be written as

\begin{eqnarray}
\nonumber
\Delta v_1 (\Delta m = 0,\, \Delta q = 0,\, \Delta A = 0) = ~~~~~~~~~\\  v_1[\p] + v_1[\de] - v_1[\hethree].
\label{delv1_eq_csr1}
\end{eqnarray}

Different terms in Eqs.\,(\ref{eq_csr1}) and (\ref{delv1_eq_csr1}) should be evaluated in a common region of rapidity $y_{\rm min}\le y \le y_{\rm max}$ and transverse momentum per constituent nucleon $(p_T/A)_{\rm min} \le p_T/A \le (p_T/A)_{\rm max}$. A common $y-p_T/A$ region is required if the coalescence mechanism is applicable. In other words, if one measures $v_1$ of $p$, $d$ and $^3$He in $y_{\rm min}\le y \le y_{\rm max}$ as a function of transverse momentum $p_T^{p},~p_T^{d}$\, and $p_T^{^3\rm{He}}$, respectively, then Eq.\,(\ref{delv1_eq_csr1}) should be evaluated in the kinematic region where $(p_T/A)_{\rm min}< (p_T^{p}),\, (p_T^{d}/2),\, (p_T^{\hethree}/3) <(p_T/A)_{\rm max}$.

\begin{table}[tbh]
\renewcommand{\arraystretch}{1.5}
\begin{tabular}{|l l l|}
\hline
Index & & ~~~~~~~~~~ $\Delta v_1$ combination  \\ 
\hline
1 && ${\p + \de} - \hethree$ \\ 
\hline
2 && ${\p + \tr} - \hefour$ \\ 
\hline
3 && $\de - \frac{1}{2}\hefour$ \\
\hline
4 && $\de + \hethree - \p - \hefour$  \\
\hline
5 && $\tr + \hethree - \de - \hefour$ \\
\hline
6 && $\hyphthree - \de - \lam $ \\
\hline
7 && $\hyphfour - \tr - \lam $ \\
\hline
\end{tabular}


\caption{Differences between the combinations formed from various light nuclei and hypernuclei. Each index represents a difference of two combinations with identical constituents, i.e., for all cases, the constituent-level mass difference is $\Delta m = 0$, the charge difference is $\Delta q = 0$, and the mass number difference is $\Delta A = 0$. Not all indices shown here are linearly independent. A set of linearly independent combinations can be found using linear algebra; one possible such set is 1, 2, 3, 6 and 7.}
\label{tab:csr}
\end{table}

Similar to the combinations in Eq.\,(\ref{delv1_eq_csr1}), various combinations are arranged in Table~\ref{tab:csr} where each index represents a difference between two combinations having identical constituents. The sum rule can be investigated experimentally in a model-independent way by each index as shown in Table~\ref{tab:csr}. The result given by any of the indices can be cross-checked by other indices. Index 1\,-\,5 are constructed from light nuclei only whereas index 6 and index 7 contain hypernuclei ($\mathrm{_{\Lambda}^{3}H}$ and $\mathrm{_{\Lambda}^{4}H}$) along with non-strange light nuclei. In indices 6 and 7, the $\Lambda$ hyperon is balanced in such a way that the net strangeness, $\Delta S$, is also zero. It is indeed very interesting to investigate the charge and strangeness dependence of the sum rule by constructing similar combinations having same or similar mass at the constituent level but different electric charge and strangeness. However this is beyond scope of this paper.

The proposed experimental test of the sum rule for light nuclei and hypernuclei can be applied to a variety of collision systems at a wide range of collision energies.  The sum rule test can also be applied to other flow harmonics, like $v_2$. It is to be noted here that the present method tests the simplified version of the sum rule where the light nucleus or hypernucleus $v_1$ is the simple addition of its constituents $v_1$. However, corrections for higher-order terms in the sum rule might be important when the $v_1$ magnitude is sufficiently larger. One should keep in mind that the higher-order terms contain $v_1(n)$ which cannot be measured in experiment. The higher-order corrections are not included in the present work.

Section~\ref{results} applies the proposed method to light nuclei in $\sqrt{s_{NN}} = 3$ GeV Au+Au collisions from STAR. At present, there are no published anisotropic flow measurements for $\mathrm{_{\Lambda}^{3}H}$ and $\mathrm{_{\Lambda}^{4}H}$ in $\sqrt{s_{NN}} = 3$ GeV Au+Au collisions. Therefore, the sum rule cannot be investigated for indices 6 and 7 at this time. Indices 6 and 7 can be exploited to predict the $v_1$ of $\mathrm{_{\Lambda}^{3}H}$ and $\mathrm{_{\Lambda}^{4}H}$ in $\sqrt{s_{NN}} = 3$ GeV Au+Au collisions:

\begin{eqnarray}
v_1[^3_\Lambda \mathrm{H}(pn\Lambda) ] = \Delta v_1  +  v_1[\de] + v_1[\lam],
\label{v1_hyph3}
\end{eqnarray}

\begin{eqnarray}
v_1[\hyphfour] = \Delta v_1  +  v_1[\tr] + v_1[\lam],
\label{v1_hyph4}
\end{eqnarray}
where $\Delta v_1$ is the difference in $v_1$ between identical constituent combinations. The $\Delta v_1$ is the measure of the sum rule check. In ideal scenario where the sum rule holds, $\Delta v_1$ should be zero. In this case, a global $\Delta v_1$ is obtained by fitting the $\Delta v_1$ calculations from other indices of Table~\ref{tab:csr}. One should fit the $\Delta v_1$ from a set of independent indices only. The indices of Table~\ref{tab:csr} are not all linearly independent. A set of linearly independent indices can be found by employing linear algebra as discussed in the next subsection.

\begin{table*}[tbh]
\renewcommand{\arraystretch}{1.5}
\begin{tabular}{|l l l l l|}
\hline
Index & & ~~~~~ $\Delta v_1$ combination     & & ~~~~~~~~~~~~~Vector  \\ 
\hline
1 && ${\p + \de} - \hethree$ & & 
$v_1=\{$1,~1,~0,~--1,~0,~0,~0,~0$\}$ \\ 
\hline
2 && ${\p + \tr} - \hefour$ & & 
$v_2=\{$1,~0,~1,~0,~--1,~0,~0,~0$\}$ \\ 
\hline
3 && $\de - \frac{1}{2}\hefour$ && 
$v_3=\{$0,~1,~0,~0,~--1/2,~0,~0,~0$\}$ \\
\hline
4 && $\de + \hethree - \p - \hefour$  && 
$v_4=\{$--1,~1,~0,~1,~--1,~0,~0,~0$\}$ \\
\hline
5 && $\tr + \hethree - \de - \hefour$ && 
$v_5=\{$0,~--1,~1,~1,~--1,~0,~0,~0$\}$ \\
\hline
6 && $\hyphthree - \de - \lam $ && 
$v_6=\{$0,~1,~0,~0,~0,~1,~--1,~0$\}$ \\
\hline
7 && $\hyphfour - \tr - \lam $ && 
$v_7=\{$0,~0,~1,~0,~0,~1,~0,~--1$\}$ \\
\hline
\end{tabular}
\caption{The vectors constructed from each combination in Table ~\ref{tab:csr}. These vectors are formulated in the {\bf R}$^8$ vector space where the basis is formed by the light nuclei and hypernuclei discussed here, namely, $p$, $d$, $t$, ${}^{3}_{}\mathrm{He}$, ${}^{4}_{}\mathrm{He}$, $\Lambda$, ${}^{3}_{\Lambda}$H and ${}^{4}_{\Lambda}$H.}
\label{tab:vec}
\end{table*}

\subsection{Evaluation of linearly independent combinations}

This subsection is dedicated to figure out the linearly independent light nuclei and hypernuclei combinations as presented in Table~\ref{tab:csr}. There are six independent measurements ($v_1$ of $\Lambda$, $p$, $d$, $t$, $^3$He, and $^4$He), using them seven combinations are made up (see Table~\ref{tab:csr}), and hence each combination must not be independent. A set of linearly independent combinations is necessary to estimate the global $\Delta v_1$ which is useful to get an overall estimation of the sum rule test and predict $v_1$ of $\mathrm{_{\Lambda}^{3}H}$ and $\mathrm{_{\Lambda}^{4}H}$ (see Eqs.~\ref{v1_hyph3},~\ref{v1_hyph4}). The global $\Delta v_1$ can be obtained by fitting the $\Delta v_1$ measurements of the independent combinations. To make the fit reliable, one has to use the independent data points in the fitting. Because any sort of correlations among the fitted data points can make the fitting procedure biased.

To find sets of linearly independent combinations among the seven combinations, linear algebra is employed where the present problem is mapped into a linear vector space. The same method of linear algebra was used to identify independent hadron combinations in a previous work~\cite{Sheikh:2021rew}. Here, it is assumed that light nuclei and hypernuclei along with the $\Lambda$ hyperon used in this approach form a basis $B = $ $\{$ $p$, $d$, $t$, ${}^{3}_{}\mathrm{He}$, ${}^{4}_{}\mathrm{He}$, $\Lambda$, ${}^{3}_{\Lambda}$H, ${}^{4}_{\Lambda}$H $\}$ of a 8-dimensional vector space, {\bf R}$^8$, where the elements of $B$ are called basis vectors in this space. This assumption is well justified since the experimental measurements of $v_1$ of each light nucleus, hypernucleus, $\Lambda$ hyperon are independent and can represent independent basis vectors of a vector space. All the combinations or indices made up from them (see Table~\ref{tab:csr}) are vectors in that space ({\bf R}$^8$), and together constitute a set of vectors, $\boldsymbol{V}=\{\boldsymbol{v}_1, \boldsymbol{v}_2,\ldots, \boldsymbol{v}_r\}$,
where $r$ is the total number of vectors in the set (in this case $r=7$).

The set of vectors, $\boldsymbol{V}=\{\boldsymbol{v}_1, \boldsymbol{v}_2,\ldots, \boldsymbol{v}_r\}$ is linearly dependent if there exists a set of non-zero scalars $(\beta_1, \beta_2, \ldots, \beta_r)$ such that
 \begin{equation}
      \sum_{i=1}^{r} \beta_i \boldsymbol{v}_i = \bold{0} 
      \label{lindep}
 \end{equation}
 where $\bold{0}$ is a null vector in the same space. In other words, the vectors are linearly dependent if at least one vector can be expressed as a linear combination of the others. The vectors in $\boldsymbol{V}$ are linearly independent when all the coefficients in Eq.\,(\ref{lindep}) are zero~\cite{riley2006mathematical}. 
 
Each vector of $\boldsymbol{V}$ can be represented as a column matrix of dimension 8$\times$1, where 8 is the dimension of the vector space in our case. This implies that Eq.~(\ref{lindep}) is a matrix equation where the seven vectors together form a matrix, $M$, of dimension 8$\times$7 and the scalars $\beta_1, \beta_2, \ldots, \beta_7$ constitute a column matrix, $B$, with dimensions 7$\times$1, i.e.,

\begin{equation}
      MB=O,
\label{lindep3}
\end{equation}
where $O$ is a null matrix of dimensions 8$\times$1. The matrix $M$ should be expressed in row-reduced echelon form by several row and column operations to solve the matrix equation, Eq.~(\ref{lindep3}). At the end, Eq.~(\ref{lindep3}) with the row-reduced form of $M$ evaluates the scalars $\beta_1, \beta_2, \ldots, \beta_7$.  

Employing the above method of linear algebra, it is found that the seven indices of Table~\ref{tab:csr} are not linearly independent. Therefore, the number of vectors in the set $\boldsymbol{V}$ can be reduced repeatedly until an independent vector subset is identified. The five indices 1, 2, 3, 6 and 7 are thus found to be linearly independent. Note that other sets of independent combinations can exist.

\section{Results and Discussions}
\label{results}

\begin{figure}
\centering
\includegraphics[width=0.5\textwidth]{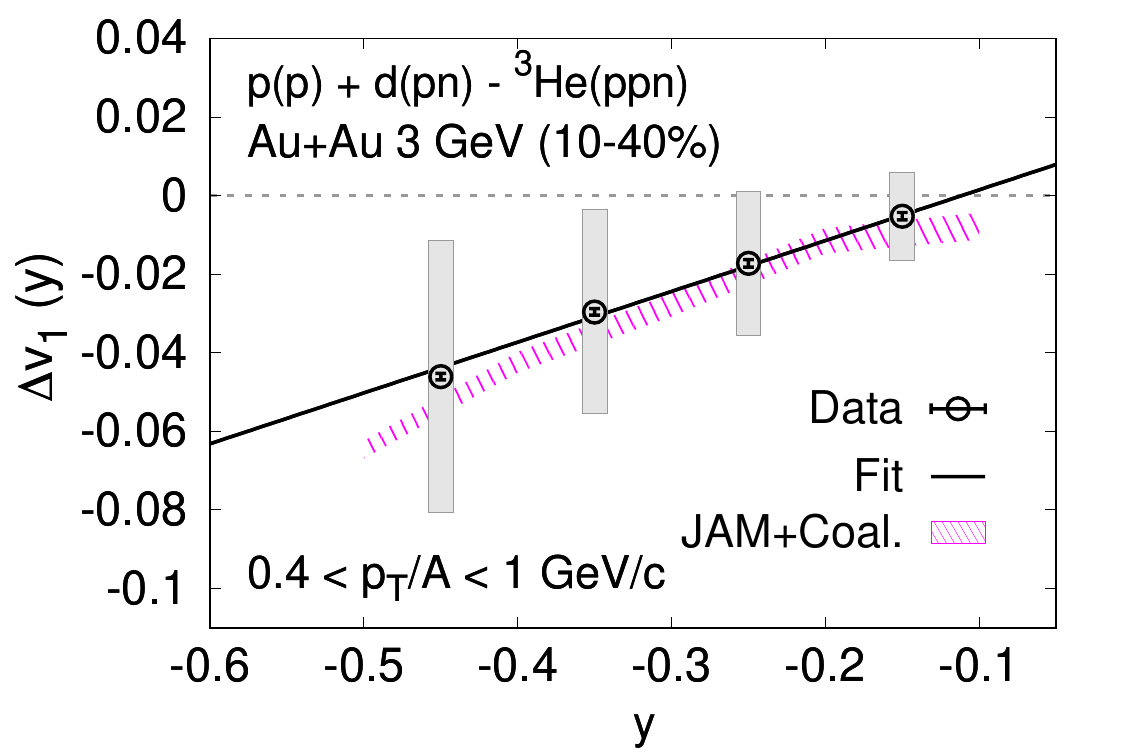}
\caption{Calculations of $\Delta v_1$ based on index 1 (see Table~\ref{tab:csr}) for $\sqrt{s_{NN}} = 3$ GeV Au+Au collisions at 10\,-\,40\% centrality, where the $v_1$ of $\hethree$ is subtracted from the combined $v_1$ of $\p$ and $\de$ as shown in Eq.~(\ref{delv1_eq_csr1}). $\p+\de$ has the same nucleon content as $\hethree$. Experimental measurements as well as the JAM (mean field)+coalescence calculations of $v_1$ are taken from Ref.~\cite{STAR:2021ozh}.} 
\label{fig:delv1_index1}
\end{figure}

\begin{figure*}
\centering
\includegraphics[width=0.495\textwidth]{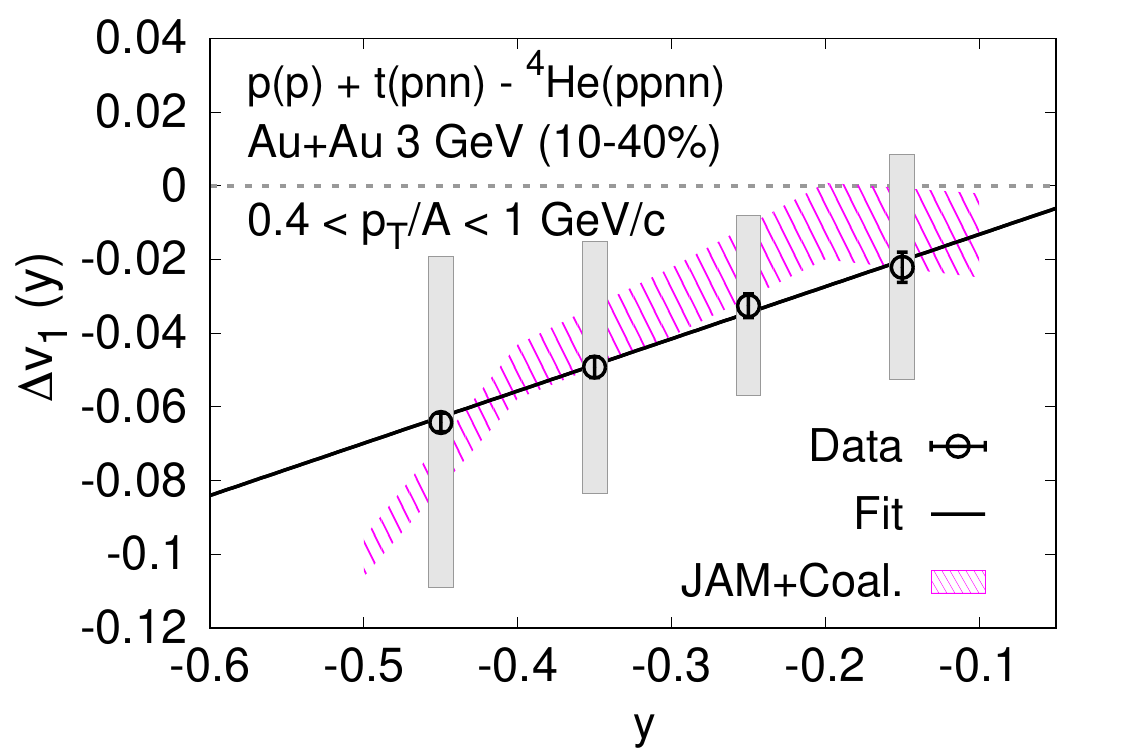}
\includegraphics[width=0.495\textwidth]{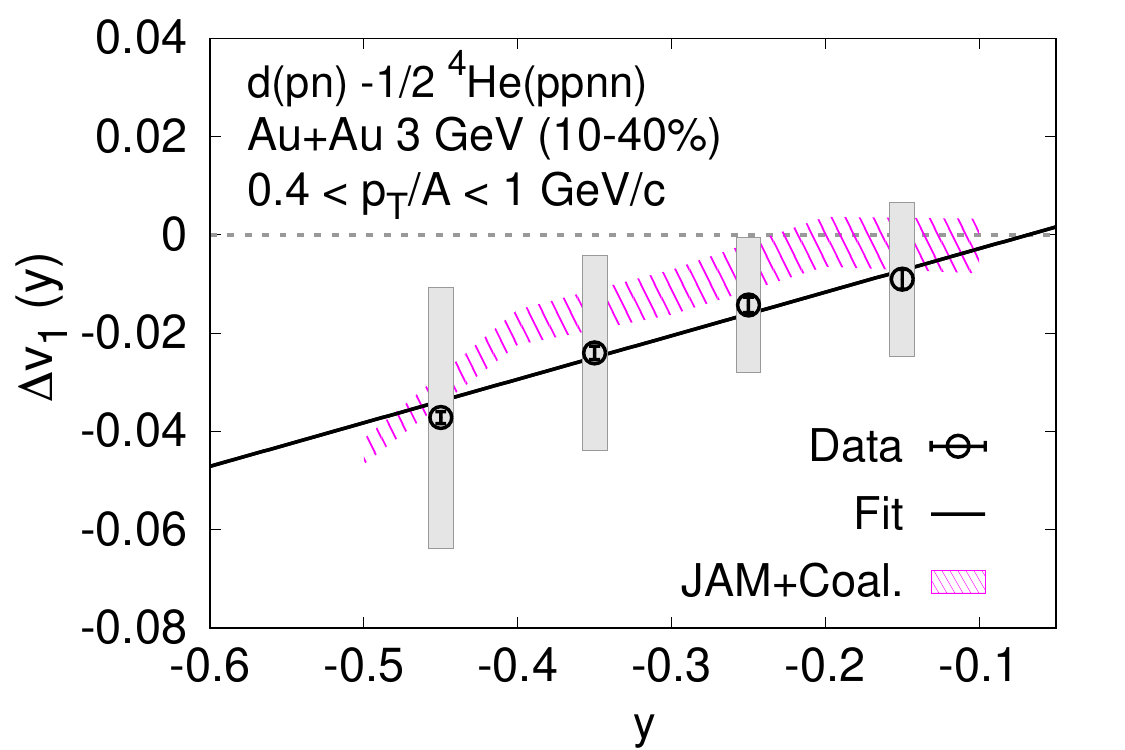}
\caption{Calculations of $\Delta v_1$ based on index 2 (left plot) and index 3 (right plot) (see Table~\ref{tab:csr}) for $\sqrt{s_{NN}} = 3$ GeV Au+Au collisions at 10\,-\,40\% centrality. The experimental measurements and the JAM (mean field)+coalescence calculations of $v_1$ for each light nucleus are taken from Ref.~\cite{STAR:2021ozh}.} 
\label{fig:delv1_index23}
\end{figure*}

The proposed method to test the coalescence sum rule for light nuclei is applied to the STAR experimental data for $\sqrt{s_{NN}} = 3$ GeV Au+Au collisions. The JAM model with baryonic mean field plus nucleon coalescence calculations are quite successful in describing the measured $v_1$ and $v_2$ for light nuclei from $\sqrt{s_{NN}} = 3$ GeV Au+Au collisions~\cite{STAR:2021ozh}.  Therefore, the findings obtained from the experimental data and the model are expected to be consistent. The JAM model + coalescence can provide further understanding of light nucleus formation, the coalescence-inspired sum rule, and scaling behavior. 
The JAM model simulates nucleon production from the initial collision phase to the final hadron transport in Au+Au collisions. 
In the mean-field mode of this model~\cite{Isse:2005nk}, nucleon evolution is performed by using a momentum-dependent potential with the incompressibility parameter, $\kappa = 380$ MeV. To simulate light nuclei, the JAM mean-field mode employs a coalescence afterburner at a fixed time of 50 fm/$c$. Each nucleon pair is boosted to the rest frame, then the relative position ($\Delta r$) and relative momentum ($\Delta p$) determines whether a light nucleus is formed. For example, if $\Delta r < 4$ fm and $\Delta p < 0.3$ GeV/$c$, then the nucleon pair is tagged as a $\de$ \cite{Sombun:2018yqh}. Other light nuclei with $A > 2$, like $\tr$, $\hethree$ and $\hefour$, are formed by adding up the constituent nucleons one by one
as per the $\Delta r$ and $\Delta p$ values in the rest frame. For more details of the model calculations, see Ref.~\cite{STAR:2021ozh}. 

Figure~\ref{fig:delv1_index1} presents estimates of $\Delta v_1$ (Eq. (\ref{delv1_eq_csr1})) as a function of rapidity, $y$, for $\sqrt{s_{NN}} = 3$ GeV Au+Au collisions at 10\,-\,40\% centrality. $\Delta v_1$ is calculated by subtracting the $v_1$ of $\hethree$ from the combined $v_1$ of $\p$ and $\de$ as described by index 1 in Table~\ref{tab:csr}.  The calculations are performed in a common region of $y-p_T/A$ space, $-0.5 < y < 0$ and $0.4 < p_T/A < 1$ GeV/$c$, using the $v_1$ measurements for light nuclei reported by STAR~\cite{STAR:2021ozh}. Calculations from JAM mean field with coalescence are also shown here. Especially near mid-rapidity, $-0.3 < y < 0$, $\Delta v_1$ is roughly consistent with zero within the measured error bars, indicating that the sum rule is followed approximately. However, moving away from mid-rapidity ($y=0$), $\Delta v_1$ magnitudes increase gradually and deviate from zero. This implies a sum rule violation which is more prominent at larger rapidity magnitudes ($y < -0.3$). 
The JAM mean field with coalescence calculations agree with the data-driven calculations within uncertainties and hence exhibit a similar violation of the sum rule.

\begin{figure*}
\centering
\includegraphics[width=0.495\textwidth]{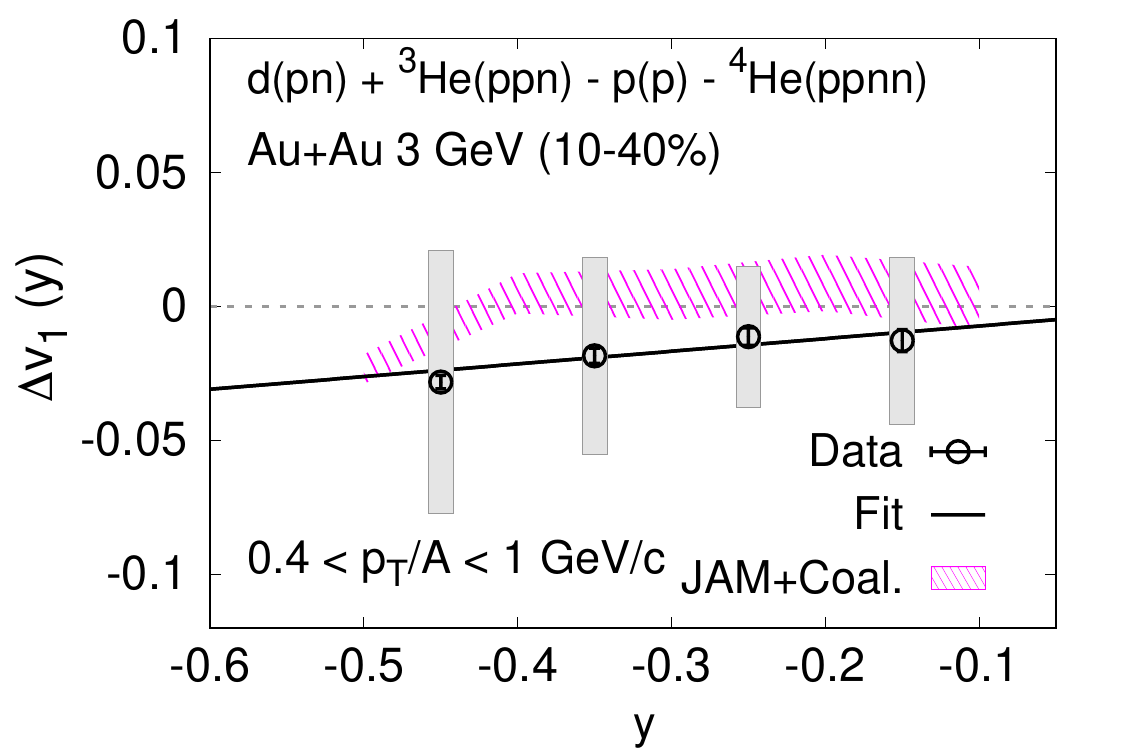}
\includegraphics[width=0.495\textwidth]{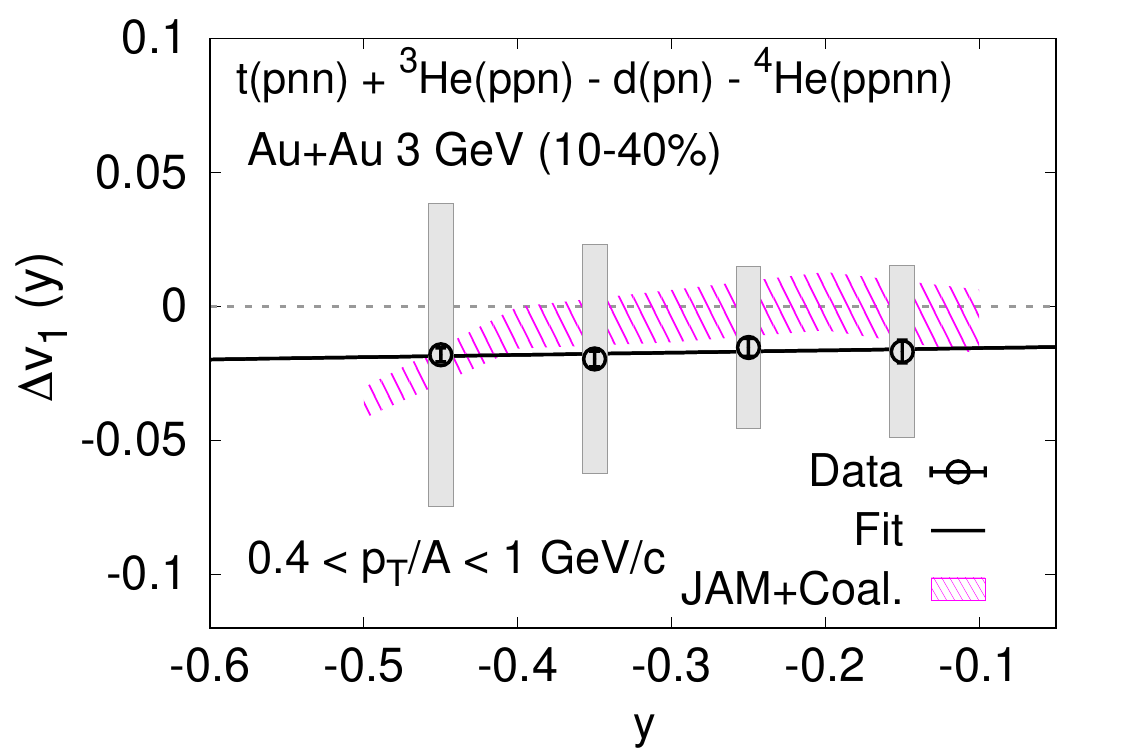}
\caption{$\Delta v_1$ based on index 4 (left plot) and index 5 (right plot) (see  Table~\ref{tab:csr}) for $\sqrt{s_{NN}} = 3$ GeV Au+Au collisions at 10\,-\,40\% centrality. The experimental measurements and the JAM (mean field)+coalescence calculations of $v_1$ for each light nucleus are taken from Ref.~\cite{STAR:2021ozh}.} 
\label{fig:delv1_index45}
\end{figure*}

Figure~\ref{fig:delv1_index23} presents $\Delta v_1$ as a function of $y$ for index 2 and index 3 of Table~\ref{tab:csr}, based on the STAR measurements~\cite{STAR:2021ozh} for $v_1$ of light nuclei at $0.4 < p_T/A < 1$ GeV/$c$ in $\sqrt{s_{NN}} = 3$ GeV Au+Au collisions at 10\,-\,40\% centrality. It is observed that $\Delta v_1$ is again consistent with zero near mid-rapidity, $-0.3 < y < 0$, within the measured uncertainties for both indices. The magnitudes of $\Delta v_1$ increase as the rapidity magnitude increases, and show a significant deviation from zero, in particular at larger rapidity magnitudes ($y < -0.3$). Nevertheless, the experimental errors are quite large, especially away from mid-rapidity. The JAM mean field with coalescence calculations are consistent with the experimental data within uncertainties. The systematic deviation of $\Delta v_1$ from zero at large rapidity magnitude suggests a breakdown of the sum rule. Recently, the STAR collaboration has found that $v_1/A$ for all light nuclei, including protons, approximately follows $A$ scaling near mid-rapidity, $-0.3 < y < 0$, and the scaling behavior worsens at $-0.4 < y < -0.3$~\cite{STAR:2021ozh}. The model calculations are consistent with these findings from STAR.

\begin{figure*}
\centering
\includegraphics[width=0.495\textwidth]{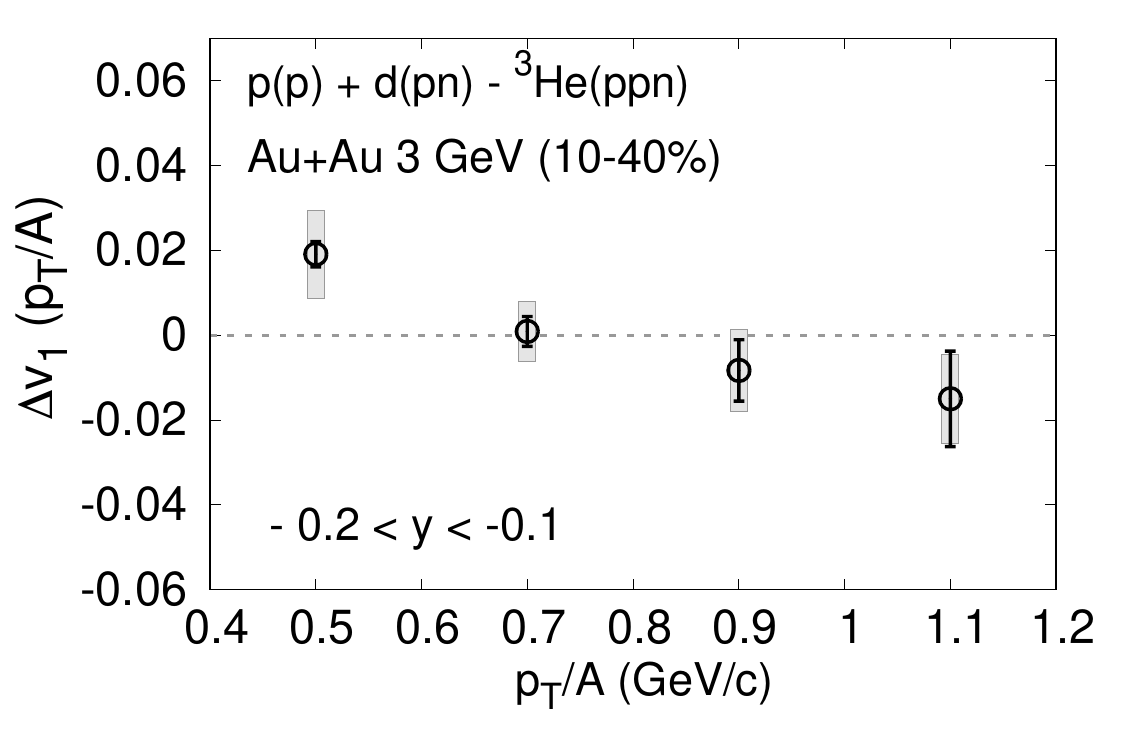}
\includegraphics[width=0.495\textwidth]{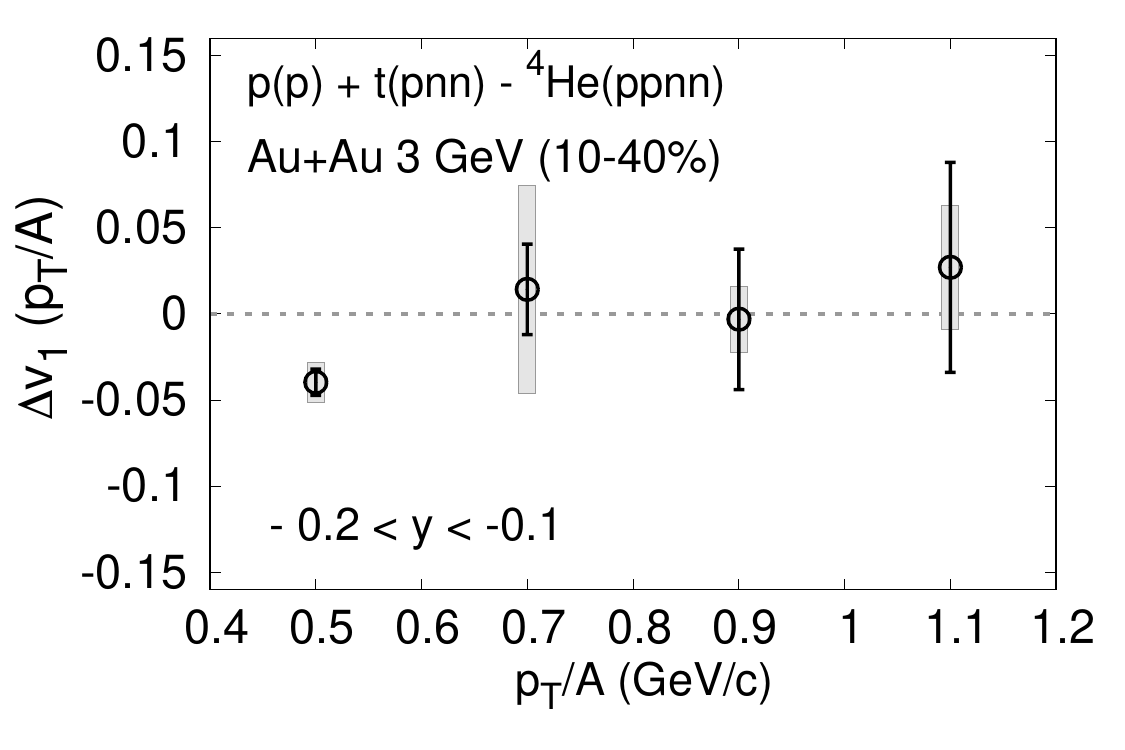}
\includegraphics[width=0.495\textwidth]{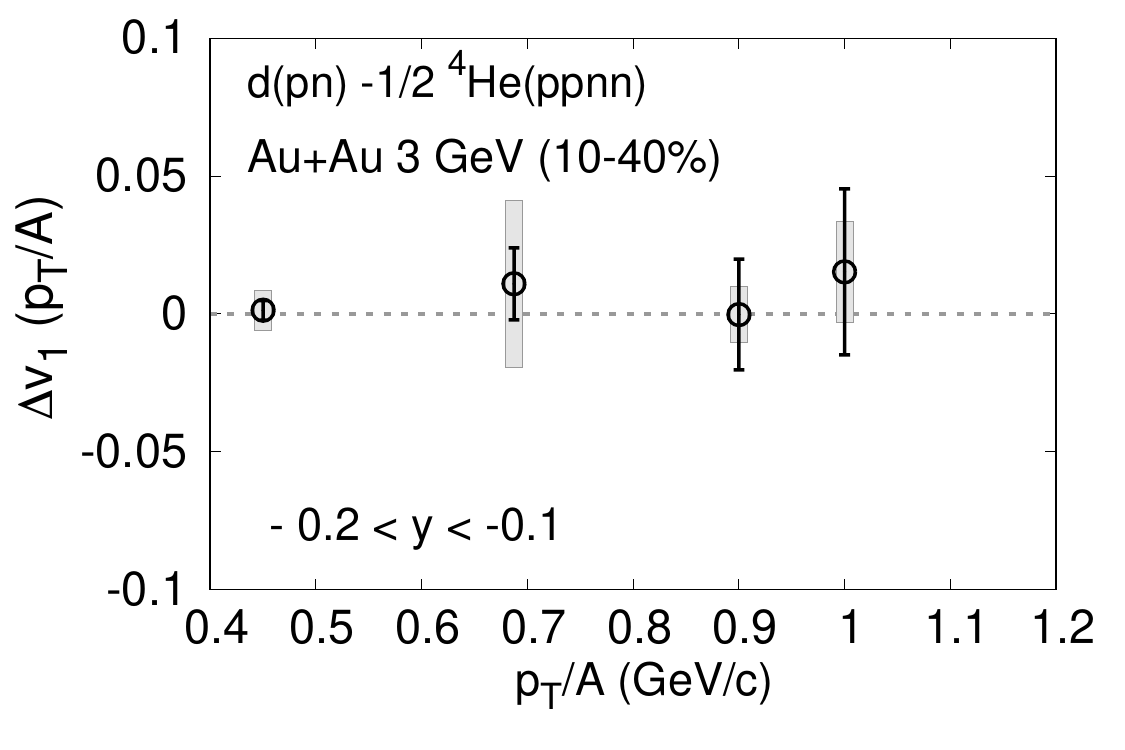}
\includegraphics[width=0.495\textwidth]{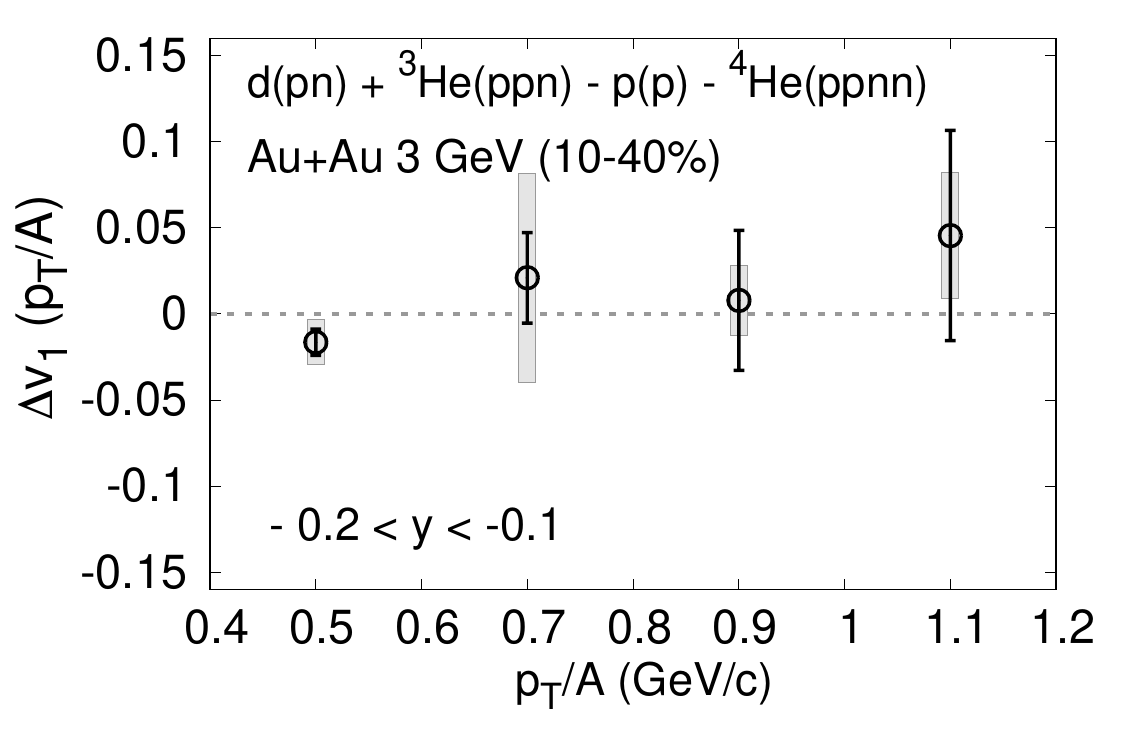}
\includegraphics[width=0.495\textwidth]{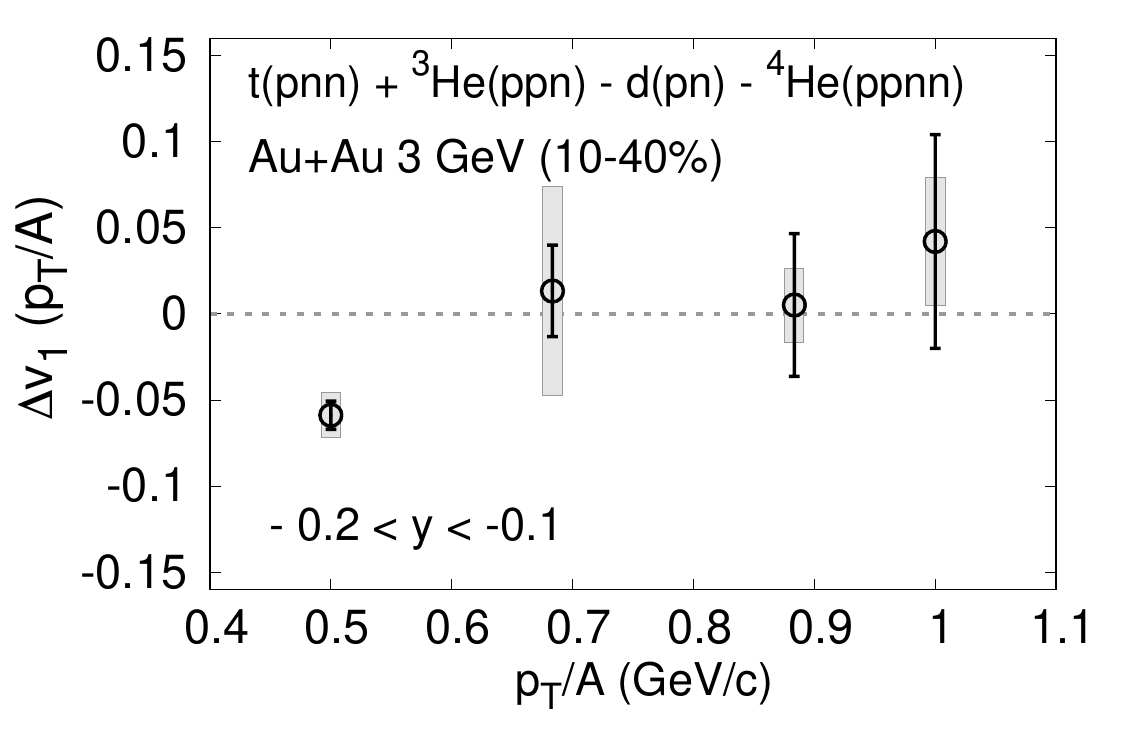}
\caption{$\Delta v_1$ as a function of $p_T/A$ for indices 1 - 5 of Table~\ref{tab:csr} in $\sqrt{s_{NN}} = 3$ GeV Au+Au collisions at 10\,-\,40\% centrality. The calculations are made in the rapidity region, $-0.2<y<-0.1$. The experimental measurements $v_1$ for each light nucleus are taken from Ref.~\cite{STAR:2021ozh}.} 
\label{fig:delv1_ptA}
\end{figure*}

\begin{figure}
\centering
\includegraphics[width=0.5\textwidth]{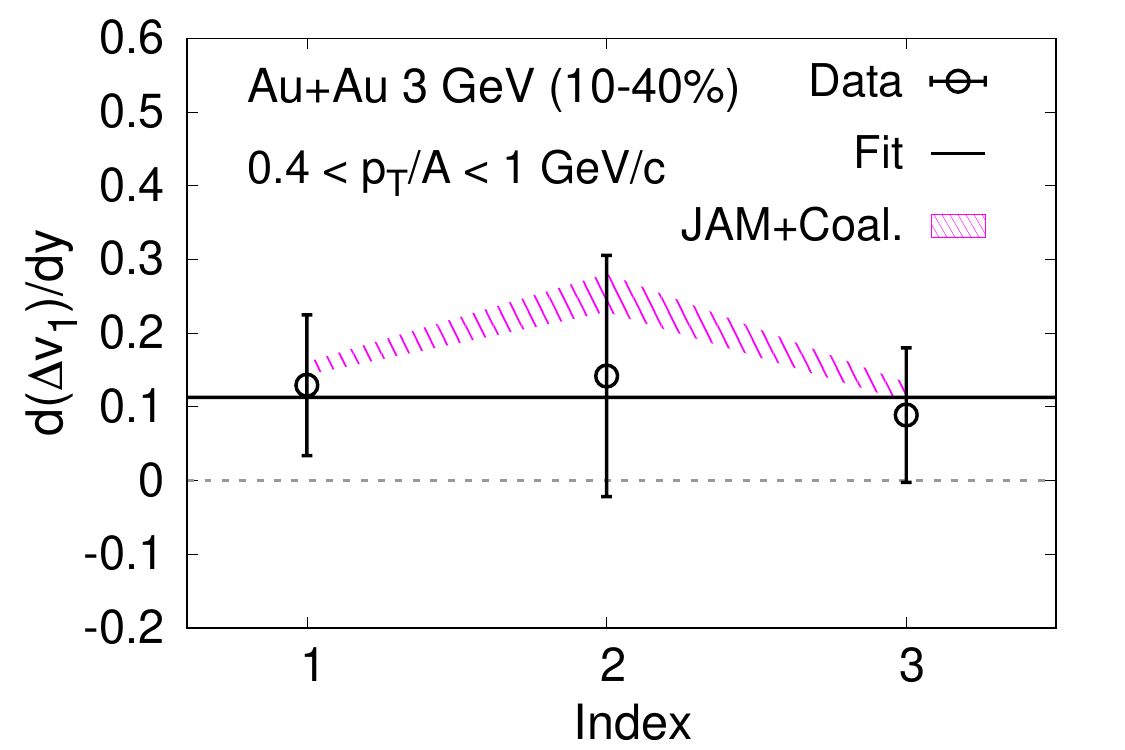}
\caption{ Estimated $\Delta v_1$ slope ($d\Delta v_1/dy$) for indices 1, 2 and 3 (see Table~\ref{tab:csr}) in Au+Au collisions at $\sqrt{s_{NN}} = 3$ GeV for 10\,-\,40\% centrality.} 
\label{fig:slope_index}
\end{figure}

Indices 4 and 5 of Table~\ref{tab:csr} have also been evaluated as a function of $y$ in 10\,-\,40\% centrality Au+Au collisions at $\sqrt{s_{NN}} = 3$ GeV, as presented in Fig.~\ref{fig:delv1_index45}. All data correspond to the same $y-p_T/A$ region: $-0.5 < y < 0$ and $0.4 < p_T/A < 1$ GeV/$c$. It is seen here that $\Delta v_1$ is close to zero within errors for all rapidity bins. The current calculations have quite a large uncertainty, particularly at the larger rapidity magnitudes. 

Exploration of all the indices of Table~\ref{tab:csr} in $p_T$ space is very interesting. Figure~\ref{fig:delv1_ptA} shows the $p_T/A$ dependence of $\Delta v_1$ for indices 1--5 in 10\,-\,40\% centrality Au+Au collisions at $\sqrt{s_{NN}} = 3$ GeV.  The calculations are made in the rapidity region, $-0.2<y<-0.1$. The calculated $\Delta v_1$ with $p_T/A$ is close to zero within the available experimental uncertainties. Nevertheless, the data point in $0.4<p_T/A<0.6$ GeV/c is a little away from zero for Index 2 and 5, and it requires further attention. It is clear that the sum rule for light nuclei is approximately valid near mid-rapidity when investigated in $p_T$ space as well.

Magnitudes of $v_1$ become larger at larger rapidity magnitudes. Beam fragments from the target rapidity region ($y < -1.045$, for $\sqrt{s_{NN}} = 3$ GeV Au+Au collisions) can be transported to the hot collision zone and the produced medium might be contaminated. Fragment contamination increases at larger rapidity magnitudes and plays a role in determining the flow of produced light nuclei. Since the fragments suffer hard interactions and more of them while being transported to the collision zone, they have different $v_1$ than a nucleon produced in the collision. The fragment contribution to light nuclei formation is likely to be greater in the region of larger rapidity magnitude and hence a simple coalescence-inspired sum rule might be less valid there.

Only $\Delta v_1$ for index 1\,-\,5 of Table~\ref{tab:csr} have been discussed so far. The other two indices (index 6 and 7) contain hydrogen hypernuclei ($\mathrm{_{\Lambda}^{3}H}$ and $\mathrm{_{\Lambda}^{4}H}$) and $v_1$ measurements for these species are not yet available. 
Based on index 6 and 7, the $v_1$ for $\mathrm{_{\Lambda}^{3}H}$ and $\mathrm{_{\Lambda}^{4}H}$ in $\sqrt{s_{NN}} = 3$ GeV Au+Au collisions are predicted in the present work.

\begin{figure*}
\centering
\includegraphics[width=0.49\textwidth]{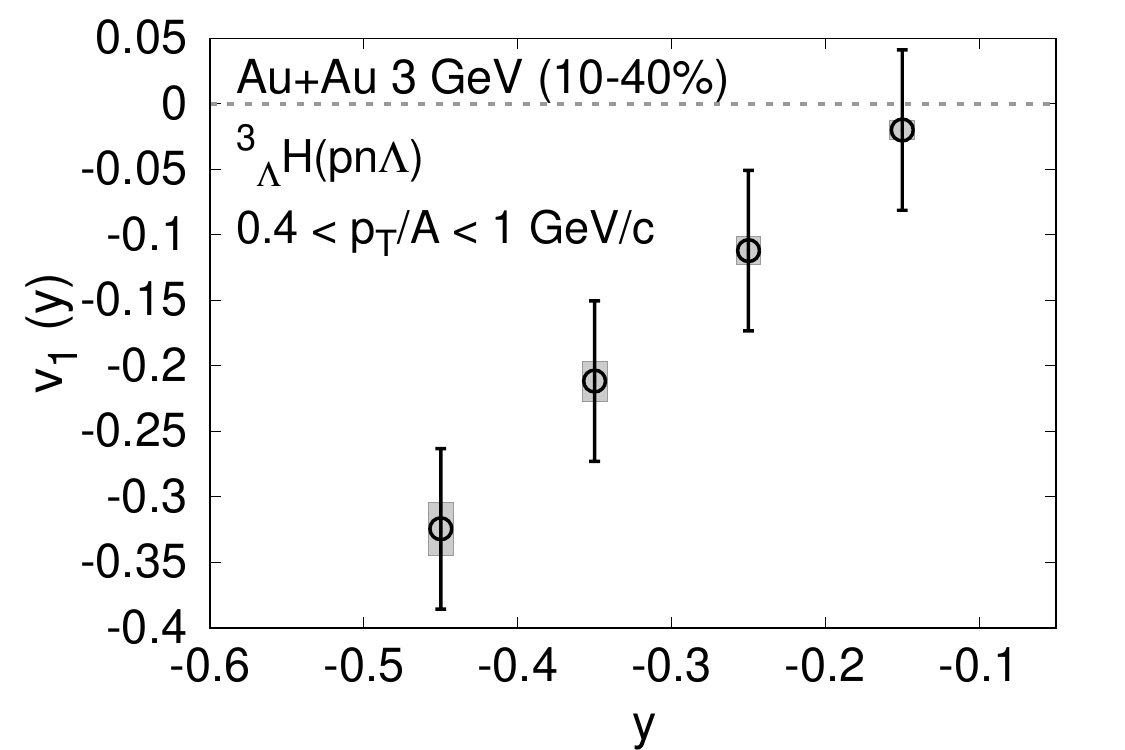}
\includegraphics[width=0.49\textwidth]{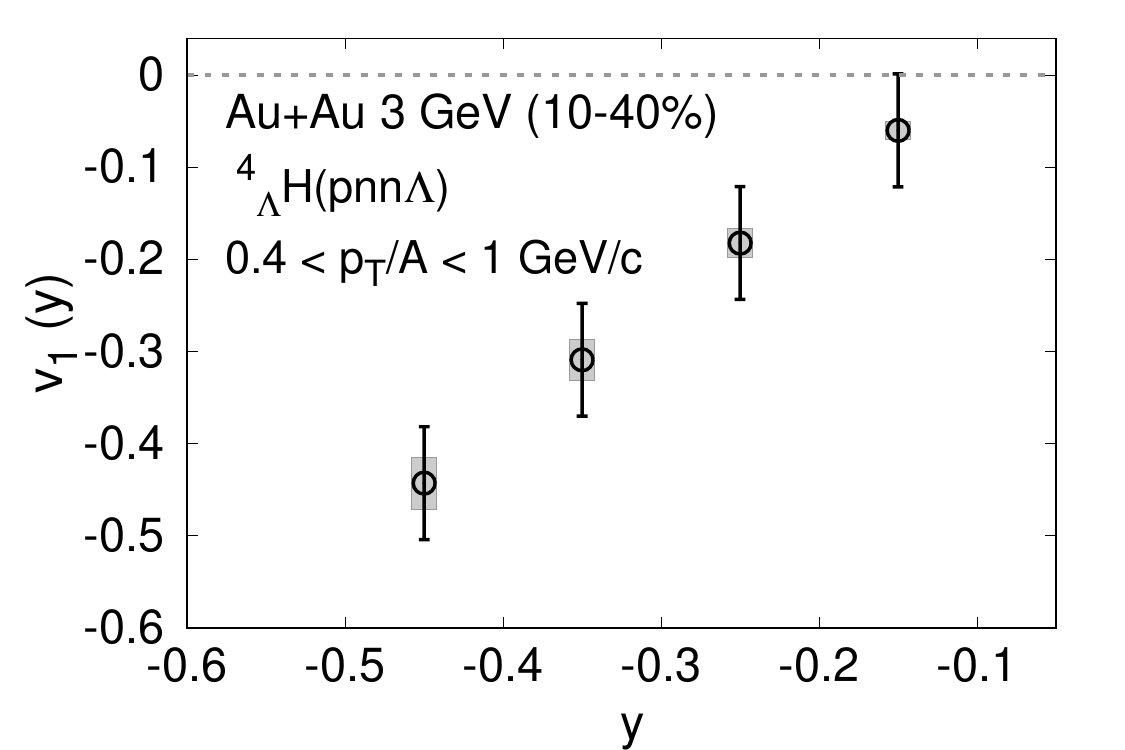}
\caption{Data-driven predictions of $v_1$ of hypernuclei $\mathrm{_{\Lambda}^{3}H}$ (left plot) and $\mathrm{_{\Lambda}^{4}H}$ (right plot) for $\sqrt{s_{NN}} = 3$ GeV Au+Au collisions at 10\,-\,40\% centrality. The predictions use index 6 and 7 of Table~\ref{tab:csr}, with the $v_1$ values taken from STAR measurements~\cite{STAR:2021ozh}.} 
\label{fig:v1_hyph}
\end{figure*}

\begin{figure}
\centering
\includegraphics[width=0.5\textwidth]{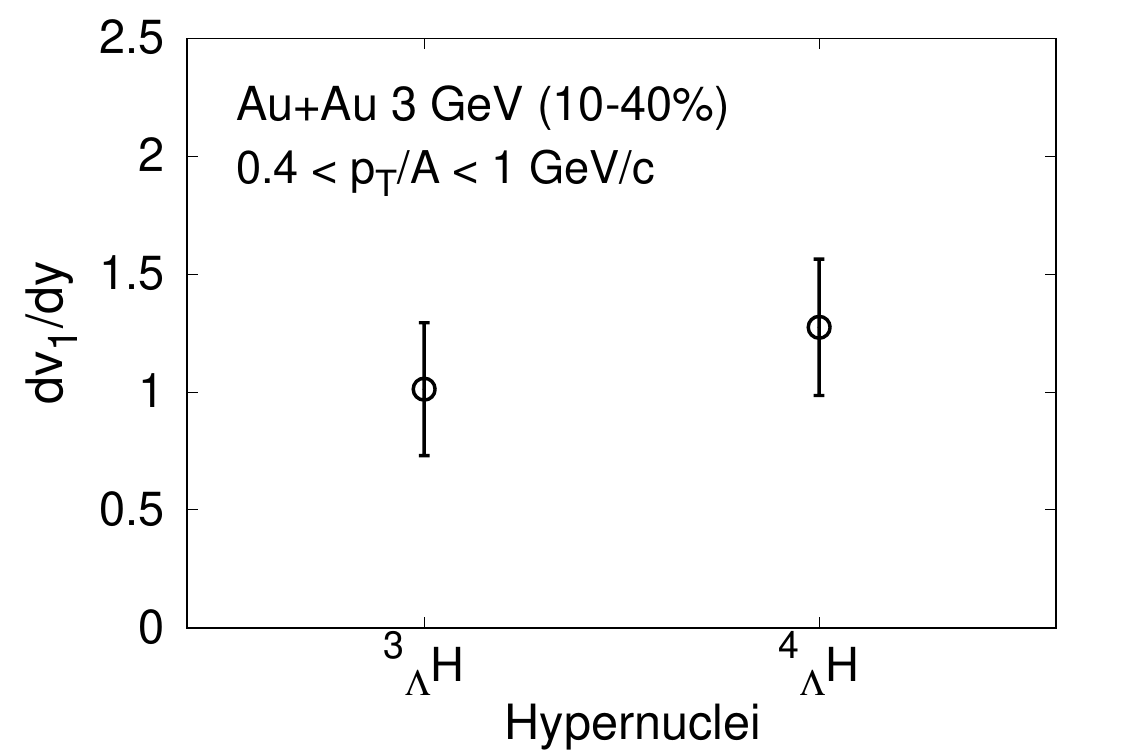}
\caption{Data-driven predictions of $v_1$ slope ($dv_1/dy$) for the hypernuclei  $\mathrm{_{\Lambda}^{3}H}$ and $\mathrm{_{\Lambda}^{4}H}$ from $\sqrt{s_{NN}} = 3$ GeV Au+Au collisions at 10\,-\,40\% centrality.} 
\label{fig:slope_hyph}
\end{figure}

The slopes $\Delta v_1(y)$ ($d\Delta v_1/dy$) for indices 1\,-\,5 of Table~\ref{tab:csr} are evaluated for Au+Au collisions at $\sqrt{s_{NN}} = 3$ GeV at 10\,-\,40\% centrality. In the ideal case where the sum rule holds, $d\Delta v_1/dy$ for all indices should be zero. Hence, the deviation of $d\Delta v_1/dy$ from zero is a measure of the sum rule violation. In Fig.~\ref{fig:slope_index}, $d\Delta v_1/dy$ is shown for three linearly independent indices, namely 1, 2 and 3 of Table~\ref{tab:csr}. 
The JAM mean field with coalescence calculations also show similar behavior. The reason to report $d\Delta v_1/dy$ only for independent indices is to fit the independent points to extract a global trend of the deviation of the calculated $\Delta v_1$ slope from zero. A constant fit of $d\Delta v_1/dy$ in Fig.~\ref{fig:slope_index} yields $C=0.15 \pm 0.007$. This is an overall measure of the sum rule violation, and has been taken into account in the present data-driven prediction of $v_1$ for hypernuclei $\mathrm{_{\Lambda}^{3}H}$ and $\mathrm{_{\Lambda}^{4}H}$ from $\sqrt{s_{NN}} = 3$ GeV Au+Au collisions.

Figure~\ref{fig:v1_hyph} reports predictions of $v_1$ for hypernuclei $\mathrm{_{\Lambda}^{3}H}$ and $\mathrm{_{\Lambda}^{4}H}$ in the reduced transverse momentum range $0.4 < p_T/A < 1$ GeV/$c$ in $\sqrt{s_{NN}} = 3$ GeV Au+Au collisions at 10\,-\,40\% centrality. The predictions are derived from Eqs.~\ref{v1_hyph3} and \ref{v1_hyph4} in a model-independent way, i.e., the terms in these equations are taken from STAR measurements~\cite{STAR:2021ozh,STAR:2021yiu}. 

The predicted $v_1$ slope ($dv_1/dy$) for $\mathrm{_{\Lambda}^{3}H}$ and $\mathrm{_{\Lambda}^{4}H}$ at $0.4 < p_T/A < 1$ GeV/$c$ from $\sqrt{s_{NN}} = 3$ GeV 10\,-\,40\% central Au+Au collisions is reported in Fig.~\ref{fig:slope_hyph}. The $v_1$ slopes are obtained by fitting the data-driven results for $\mathrm{_{\Lambda}^{3}H}$ and $\mathrm{_{\Lambda}^{4}H}$ as shown in Fig.~\ref{fig:v1_hyph}. The extracted slope values are
$dv_1/dy$ $(\mathrm{_{\Lambda}^{3}H}) = 1.012 \pm 0.282$ and $dv_1/dy$ $(\mathrm{_{\Lambda}^{4}H}) = 1.274 \pm 0.289$. The STAR collaboration has already collected large data samples that will provide greatly increased statistics for hypernuclei. Current predictions of $v_1$ and $dv_1/dy$ for $\mathrm{_{\Lambda}^{3}H}$ and $\mathrm{_{\Lambda}^{4}H}$ will serve as a baseline for ongoing and future measurements.

\section{Summary}
\label{summary}
 Light nuclei and hypernuclei carry important information on the collective motion of the produced nuclear matter in heavy-ion collisions. However their production mechanism remains uncertain. Light nuclei and hypernuclei can be formed by coalescence of nucleons and $\Lambda$-hyperons which are close to each other in both coordinate and momentum space. 
 Atomic mass number scaling for light nuclei, a consequence of the coalescence mechanism, is found to hold approximately near mid-rapidity, whereas departures from this scaling behavior appear to occur, with marginal statistical significance,
 away from mid-rapidity~\cite{STAR:2021ozh}. This traditional scaling pattern involves dividing the anisotropic flow coefficients of a light nucleus or hypernucleus by its number of constituent baryons. This scaling ignores the mass and charge differences among the constituents, which can be expected to influence the coalescence mechanism. In this article, an approach is discussed to test the coalescence-inspired sum rule for light nuclei and hypernuclei in a data-driven way, where each constituent is balanced appropriately in terms of mass and charge. In this approach, various light nuclei and hypernuclei are combined, and then the combinations having identical constituents are compared, i.e., comparisons are made for the same mass and same charge at the constituent level. The method is applied to STAR flow measurements for light nuclei from $\sqrt{s_{NN}} = 3$ GeV Au+Au collisions. It is observed that the sum rule is valid approximately near mid-rapidity, $-0.3 < y < 0$, and it is violated away from mid-rapidity, $y < -0.3$, with $1.84\sigma$ statistical significance. 
 The JAM mean-field with coalescence calculations also are consistent with the data driven results. There is an overall consistency between the calculations presented here regarding the sum rule and STAR findings on $A$ scaling for light nuclei. The $v_1$ of hypernuclei $\mathrm{_{\Lambda}^{3}H}$ and $\mathrm{_{\Lambda}^{4}H}$ is predicted in a data-driven way over a reduced transverse momentum range $0.4 < p_T/A < 1$ GeV/$c$ for $\sqrt{s_{NN}} = 3$ GeV Au+Au collisions at 10\,-\,40\% centrality. The predicted $v_1$ slope is $dv_1/dy = 1.012 \pm 0.282$ and $1.274 \pm 0.289$ for $\mathrm{_{\Lambda}^{3}H}$ and $\mathrm{_{\Lambda}^{4}H}$, respectively. The STAR collaboration has acquired large data samples that will provide greatly increased statistics for hypernuclei over a range of collision energies.  The current predictions will serve as a baseline for these upcoming hypernuclear $v_1$ measurements.

\section*{Acknowledgment}
I am thankful to Declan Keane, Prithwish Tribedy, Fuqiang Wang and Xionghong He for insightful discussions. I would also like to thank many of the STAR collaborators for the fruitful discussions. Thanks to Santosh Kumar Das for carefully reading the article. I acknowledge support from the Office of Nuclear Physics within the US DOE Office of Science, under Grant DE-FG02-89ER40531.

\bibliography{main}

\end{document}